\documentclass[a4paper,twocolumn,amsmath,amssymb,aps,prl,10pt,superscriptaddress,showpacs,groupedaddress]{revtex4-1}

\usepackage{amsmath}
\usepackage{amssymb}
\usepackage{graphicx}
\usepackage[utf8]{inputenc}
\usepackage[T1]{fontenc}
\usepackage{color}
\newcommand{\av}[1]{\langle\, #1\,\rangle}

\usepackage{braket}
\begin{document}
\title{Mediator assisted cooling in quantum annealing}
\author{M.\ Pino}
\author{Juan Jos\'e Garc\' ia-Ripoll}
\affiliation{Institute of Fundamental Physics IFF-CSIC, Calle Serrano 113b, Madrid 28006, Spain}

\begin{abstract}
We show a significant reduction of errors for an architecture of quantum annealers (QA) where bosonic modes mediate the interaction between qubits. These systems have a large redundancy in the subspace of solutions, supported by arbitrarily large bosonic occupations. We explain how this redundancy leads to a mitigation of errors when the bosonic modes operate in the ultrastrong coupling regime. Numerical simulations also predict a large increase of qubit coherence for a specific annealing problem with mediated interactions. We provide evidences that noise reduction occurs in more general types of quantum computers with similar architectures.
\end{abstract}
\maketitle

\paragraph{Introduction.}

A quantum annealer\ \cite{apolloni1989quantum,kadowaki1998quantum,somorjai1991novel} is a device that evolves adiabatically a quantum system from an easy to prepare initial ground state, to a final one that encodes the solution of a problem. An adiabatic quantum computer (AQC) is a more powerful and general device\ \cite{albash2018adiabatic}, that prepares the outcome of an arbitrary quantum computation through a similar adiabatic process. It has been argued that AQC may have some intrinsic robustness against decoherence when compared with the equivalent\ \cite{aharonov2008adiabatic, kempe2006complexity} gate-based quantum computer\ \cite{childs2001robustness}. However, the adiabatic condition demands long evolution times\ \cite{galindo2012quantum}, during which noisy devices can be excited, ruining the adiabatic computation.

There are two main strategies to reduce the effect of noise in quantum devices. In error protection schemes, the quantum register decouples from the noise by design\ \cite{kitaev2003fault,gladchenko2009superconducting, pino2015, bell2016spectroscopic,bell2014}, typically with the help of symmetries or topology. In error correcting schemes, information is stored redundantly in logical qubits\ \cite{shor1995scheme}, composed of multiple physical qubits, with protocols to detect and correct errors. These strategies have been applied to mitigate the effect of noise in AQC. There are protection schemes based on energy gaps\ \cite{jordan2006error, bookatz2015error}, dynamical decoupling\ \cite{lidar2008towards}, Zeno effect\ \cite{paz2012zeno}, or nested quantum computing\ \cite{vinci2016nested} and some error correction schemes have been proposed and tested in the D-wave QA\ \cite{pudenz2014error}. Almost all of these schemes have a considerable experimental overhead---additional qubits for redundant encoding, error detection and correction operations---that may be comparable to the resources demanded by error-corrected gate-based quantum computers\ \cite{young2013error, sarovar2013error}.

\begin{figure}[t!]
\begin{centering}
\includegraphics[width=0.75\columnwidth]{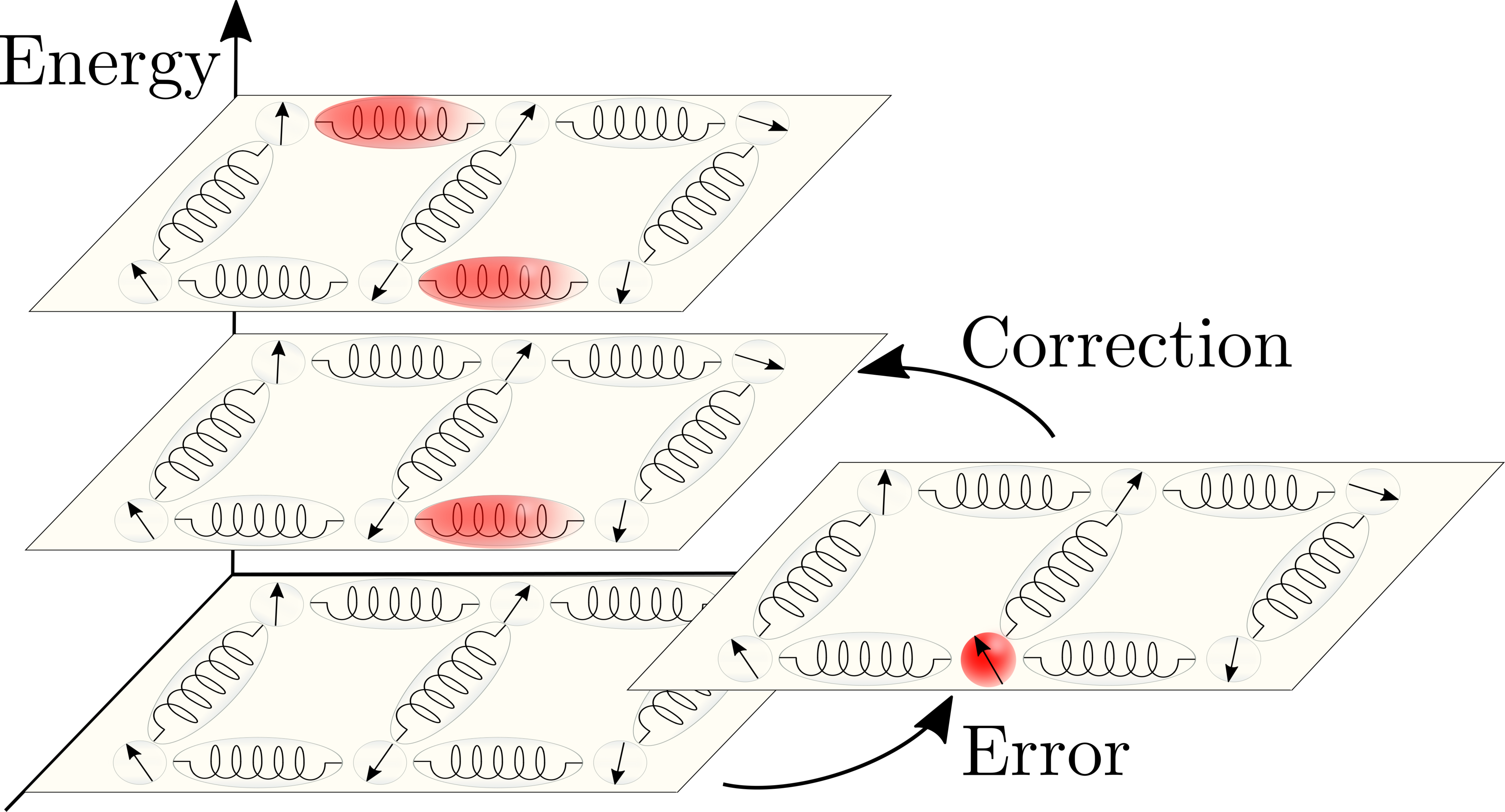}
\end{centering}
\caption{ Scheme for the Hilbert space of a  QA architecture in which the interaction between qubit (arrows) are mediated by bosonic modes (springs).  Effective qubit Hamiltonians at different energies can be defined for different bosonic configurations.  All of them become equal at the end of the annealing passage, which implies a large redundancy of the subspace of solution. This redundancy is used to attenuate the effect of noise in the ultrastrong coupling regime where qubit and bosonic excitation become close in energy. }\label{Fig:explanation}
\end{figure}

Here we show a large suppression of the effect of noise in an architecture of QA where the interactions between qubits are mediated by bosonic modes---LC resonator (or transmission line) in superconducting circuits\ \cite{majer2007coupling, allman2014tunable} and phonons in ion traps \cite{porras2004effective}. The mechanism is based on a transfer of energy and entropy from errors in the quantum register, to excitations in the bosonic degrees of freedom, see Fig.\ \ref{Fig:explanation}. This autonomous error correction is favored by the ultra-strong coupling regime of qubit-boson interactions\ \cite{yoshihara2017superconducting, forn2017ultrastrong, bernardis2018, jaako2016ultrastrong} and has no real overhead, since bosonic couplers are already present in many blueprints of quantum computers, as agents to facilitate short and long range interactions\ \cite{mukai2019superconducting,dicarlo2009demonstration,Srinivasa2016}.

\paragraph{Architecture with mediated interactions.--} 

We will compare two designs of QA's. Our reference is an Ising model with direct interactions $(\hbar=1),$
\begin{equation}
    H_{is} =  \sum_{i=1}^{L_s} \frac{h_i}{2}\sigma_i^z  + \sum_{i,j=1}^{L_s} J_{ij} \sigma_i^x\sigma_{j}^x\label{eq:is}.
\end{equation}
We can design an annealing schedule that starts with $J=0$ and ends up with $|J|\gg |h|,$ to prepare the ground state of the Ising problem. We will also consider a generalized spin-boson (SB) model with mediated couplings,  
\begin{align}
H_{sb} &=  \sum_{i=1}^{L_s} \frac{h_i}{2}\sigma_i^z + \sum_{i,r=1}^{L_s, L_b}g_{ir} \sigma_i^x(b_r - b^\dagger_{r}) +\omega \sum_i b^{\dagger}_i b_i\label{eq:sb}.
\end{align}
This model differs from the traditional Hamiltonian where modes implement a local bosonic environment\  \cite{caldeira1981influence,wubs2006gauging, arceci2017dissipative,theis2018gap,chasseur2018environmental}. Previous Hamiltonian mimics the Ising model at low energies\ \cite{kurcz2014hybrid,pino2018quantum}, using $L_b$ bosonic modes to simulate $J_{ij}\sim \sum_r g_{ir}g_{jr}/\omega.$ It is possible to engineer an annealing schedule\ \cite{pino2018quantum} for Eq.\ \eqref{eq:sb} that reproduces the outcome of Eq.\ \eqref{eq:is}. However, success in the SB annealing is more general, as we may afford having bosonic excitations that introduce no errors in the quantum register.

To be precise, let us define the effective Hamiltonians   $H^{(\textbf{n})}=\mathbb{P}_{\textbf{n}}\widetilde{H}_{sb}\mathbb{P}_{\textbf{n}}$ for any configuration of the bosonic modes $n=(n_0,n_1,\dots, n_{L_b})$, where the tilde denote that it has been transformed by the polaron unitary $\mathcal{U} = \exp{[-\sigma_i^x\sum \phi_{ir}(b^{\dagger}_r-b_r)}]$  with $\phi_{ir}=g_{ir}/\omega$\ \cite{kurcz2014hybrid}.  All of the effective models are Ising Hamiltonians Eq.\ \eqref{eq:is}, $H^{(\textbf{n})}=H_{is}(h^{(\textbf{n})},J^{(\textbf{n})})$ with renormalized parameters
\begin{align}
 J^{(\textbf{n})}_{ij} &= J^{(\textbf{0})}_{ij}, \label{eq:renor_1}\\
 h^{(\textbf{n})}_i    &=   h^{(\textbf{0})}_i\prod_{r=1}^{L_b}\ L_{n_r}\left(4\phi_{ir}^2 \right),\label{eq:renor_2}
\end{align}
where $h^{(\textbf{0})}_i=h_i e^{-2\sum_r\phi_{ir}^2}$ and $ J^{(\textbf{0})}_{ij}=\omega\sum_r \phi_{ir}\phi_{jr}$ are the low-energy effective parameters, and $L_k(x)$ are Laguerre polynomials\ \cite{cahill1857ordered}. At the end of the quantum annealing passage, $h_i=0,$ all Hamiltonians are identical and have the same spin configurations as low-energy states. Therefore, if the annealing succeeds, it can do so for many different configurations of the bosonic modes, many of which include excited sectors of the Hilbert space.

We will analyze how this redundancy in the subspace of solutions can mitigate the effects of noise. For that, we compare annealing passages in both architectures, parameterized by the relative annealing time $s=t/T.$ For direct coupling Eq.\ \eqref{eq:is}, we use a one-dimenional chain with equal fields $h_i=\omega_0(1-s)$ and interactions $J_{ij}=- \eta\delta_{i,i+1}\omega_0 s$ (ferro or antiferro $\eta=\pm1$). For mediated couplings Eq.\ \eqref{eq:sb}, we use the same number of qubits as bosonic modes $L,$ local fields are $h=\omega_0[1-\kappa(s)]$ and  couplings $g_{ir}= \sqrt{\omega_0 \omega \kappa(s)}(\delta_{i,r}+ \eta\delta_{i,r+1}).$ The ramps $\kappa(s)$ are designed to have the same ground state expected values of $C=\sum_i\av{\sigma_i\sigma_{i+1}}$ in both models at all $s$,  $\kappa(s)=C_{sb}^{-1}(C_{is}(s)).$ This condition is approximately equivalent to $h^{(\textbf{0})}(s)=\omega_0(1-s)$ and $J^{(\textbf{0})}(s)=\omega_0 s.$ The typical value of the qubit frequency is $\omega_0$ and we consider the ultrastrong coupling regime, where $\omega_0\simeq\omega.$ For all of our numerical simulations we have chosen frequencies $\omega_0=\omega.$  We note that the minimum gaps for direct and mediated couplings obey the same law $\Delta_m\sim L^{-z}$ with $z=1$, which implies the same complexity class for direct and mediated couplings\ \cite{kurcz2014hybrid}. See Supplemental Material, section A, for a detailed comparison on how the gap closes in both models.

\paragraph{Error suppression and error correction.--}
Errors in an adiabatic passage can be seen as transitions to excited states. Let us now study the dynamics of those errors for  the transitionally invariant Ising chain with nearest-neighbors connectivity in Eq.\ \eqref{eq:is}. The qubit operator $\gamma_q^\dagger$ that creates an error with momentum $q$ has an analogue in the SB that creates a pure qubit excitation $\ket{er_q}=\widetilde{\gamma}_q^\dagger\ket{\widetilde{\psi}_{gs}^{(\textbf{0})}},$ where $\ket{\widetilde{\psi}_{gs}^{(\textbf{0})}}=U^\dagger\ket{\varphi_{gs}^{(\textbf{0})}, \textbf{0}} $ is the ground state of the Ising model $H^{(\textbf{0})}$ and the bosonic vacuum transformed to the polaron basis. However, the SB model also supports bosonic excitations $\ket{es_q}=\widetilde{b}_q^\dagger\ket{\widetilde{\psi}_{gs}^{(\textbf{0})}},$ which implement \emph{excited solutions}. The dynamics in the spin sector of these solutions is given by
$H^{(q)}=H_{is}(h^{(q)},J^{(\textbf{0})})$ with  
\begin{equation}
 h^{(q)}_i    =   \frac{h_i^{(\textbf{0})}}{L} \sum_{r=1}^{L}\left[ L_1(4\phi_{ir}^2)+4\cos\left(q\right)\phi_{ir}\phi_{ir+1}\right] \label{eq:renor_q}.
\end{equation}
Both $H^{(q)}$ and $H^{(\textbf{0})}$ give rise to the same solutions along the annealing process, with the same complexity class.  When we perform the annealing passage in the ultrastrong coupling SB model, $\omega_{0}\approx \omega,$ the error states $\ket{er_q}$ and the excited solutions $\ket{es_q}$ experience avoided crossings at specific values of the dimensionless time $s_c.$ At those points, large fluctuations in the bosonic modes cannot be captured by the polaron ansatz, and the states couple with strength\ \cite{kurcz2014hybrid}
\begin{equation}
g_q=|\braket{er_q|H_{sb}|es_q}|\sim \sqrt{\frac{s\omega_0}{\omega} }(1-s)\omega_0.\label{Eq:hyb}
\end{equation}
The SB model eigenstates are actual superpositions $\ket{\Psi_\pm}=\frac{1}{\sqrt{2}}(\ket{er_q}\pm\ket{es_q}),$ which facilitate two new mechanisms that improve the annealing. First, an error created at early times $s\ll s_c$ can be transferred to an excited solution around the crossing point. This mechanism for \emph{error correction} only works when the passage is adiabatic with respect to the level crossing $T\gg 1/g_q.$ The other possibility is that the error states $\ket{\phi(t_0)}=\ket{er_q}$ dephase under the action of the effective Hamiltonian around the crossing. Initially, the reduced density matrix $\rho(t)$ of the error state in the spin sector has no overlap with the ground state manifold, that is $\mathcal{F}(t_0)=tr(\mathbb{P}_{gs}\rho(t))=0$ for the ground state projector $\mathbb{P}_{gs}.$ However, the overlap improves to around $\mathcal{F}\approx 1/2$ as the state dephases at long time $\rho(t\gg 1/g_q)=\frac{1}{\sqrt{2}}(\ket{\Psi_+}\bra{\Psi_+}+\ket{\Psi_-}\bra{\Psi_-}).$  We call this mechanism \emph{error suppression}.

\begin{figure}[t!]
\begin{centering}
\includegraphics[width=1.0\columnwidth]{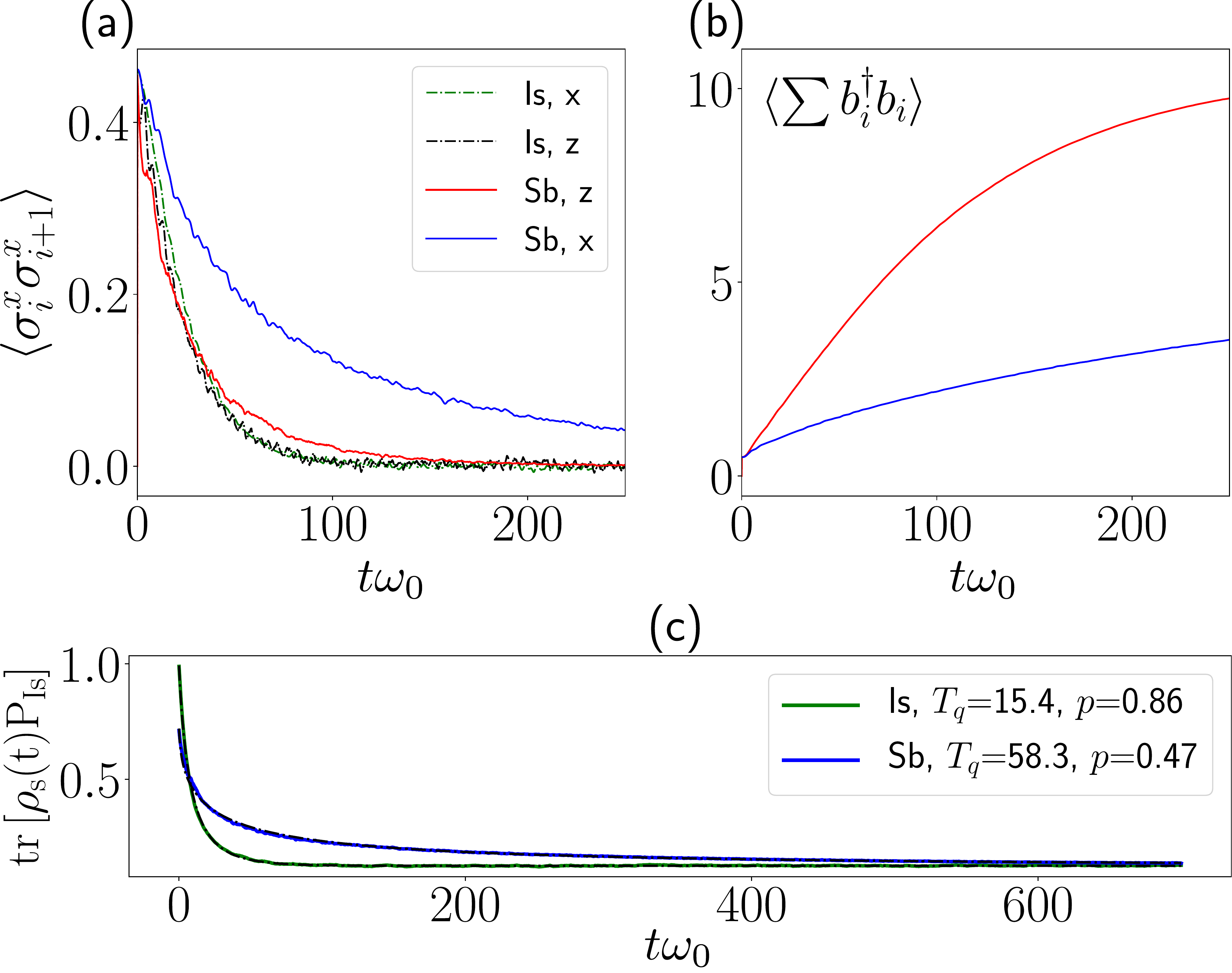}
\par\end{centering}
\vspace{0.1cm}
\caption{ Time evolution for the ferromagnetic Ising in transverse field with parameter $s=0.25$ and $L=3$ spins.  Different curves correspond to direct (Is) and mediated couplings (Sb) with noise in $x$ or $z$ direction. The power spectrum of the noise is flat with strength $\gamma=0.2.$ (a) Decay of correlations and (b) total number of bosons , $N_b=\sum_i b_i^\dagger b_i$, as a function of time.  (c) Decay of the density matrix (reduced spin density matrix in SB) fidelity with the Ising ground state as a function of time. The semi-dotted black lines are fits to  $Y(T)=\{[1+a\ \exp(-(t/T_q)^p)]/2\}^L$, which gives the value of parameters shown in the legend. A cutoff in the number of bosons per-site $n_{co}=8$ is used.
}\label{Fig:1point}
\end{figure}

\paragraph{Numerical simulations}
We have studied the one-dimensional Ising model with transverse field in a noisy environment. We add a stochastic term $1/2\sum_{i=1}^N f_i(t)\sigma_i^{\theta}$ to Eqs.\ \eqref{eq:is} and \eqref{eq:sb}\ \cite{sarovar2013error}. This is a sum of uncorrelated white Gaussian noises $f_{i}(t)$ with power spectra $S(\omega)= \frac{\gamma^2}{2\pi\omega_0}$, as described in Supplemental Material section B. The numerical integration of the resulting equations of motion have been performed with exact Lanczos methods and full wavefunctions for several realizations of the noise.  This noise couples to the spins along one of two directions $\theta=x,z$ with strengths $\gamma=0.1, 0.2.$ This extreme form of noise excites all energy scales with equal probability, heating and dephasing the spins to infinite temperature at timescales $\omega_0 T_1=\omega_0 T_2=1/(2\gamma^2)$\ \cite{paladino20141}. We assume that bosons are not affected by the noise, because resonators and cavities have a consistently larger quality factor than superconducting qubits.

We analyze the error suppression mechanism by comparing the qubit dynamics in both QA architectures, at a quarter of an adiabatic passage $s=0.25$ with $L=3$ qubits. At this point, spins and bosons are strongly hybridized. In Fig.\ \ref{Fig:1point}(a), the decay of two qubit correlations is plotted as a function of time, for the Ising and SB Hamiltonians, with noise along $x$ or $z$ direction.  The decoherence of the SB model is significantly slowed down, particularly for noise along $x.$ Panel (b) shows the total number of bosons in the hybrid model. Note how it grows rapidly for noise along z direction, indicating a stronger heating of the interaction mediators.

We characterize the error suppression for noise along the $x$ direction using the overlap with the ground state manifold $Y(t) = \mathrm{tr}(\rho(t)\mathbb{P}_{gs}),$ of the spin reduced density matrix $\rho(t)$ in both the Ising and SB passages. As shown in Fig.\ \ref{Fig:1point}(c) the hybrid SB model exhibits a slower decay, extending the lifetime of information by, at least, one order of magnitude. The data can be fitted to a law (semi-dashed black lines) that is a generalization of the decay of $L$ uncorrelated qubits
\begin{equation}
 Y(t) = \left[\frac{1+a\ e^{-(t/T_q)^p}}{2}\right]^L, \label{Eq:fit}
\end{equation}
with free parameters $a,T_q, p.$ The fits give $T_q=15.41\pm 0.06$ and $p=0.86\pm 0.02$ for Ising model, and $T_q =58.3\pm 0.1$ and $p=0.47\pm 0.01$ for SB. This is a significant noise reduction that extends the lifetime of the combined model beyond the decoherence time of non-interacting qubits, $T_1$ and $T_2.$ Note that $Y(t=0)<1$ and $a<1$ in the SB is a consequence of hybridization, but this is irrelevant because the spin sector always reproduces good normalized expectation values (Fig.\ \ref{Fig:1point} (a)).

\begin{figure}[t!]
\begin{centering}
\includegraphics[width=1.02\columnwidth]{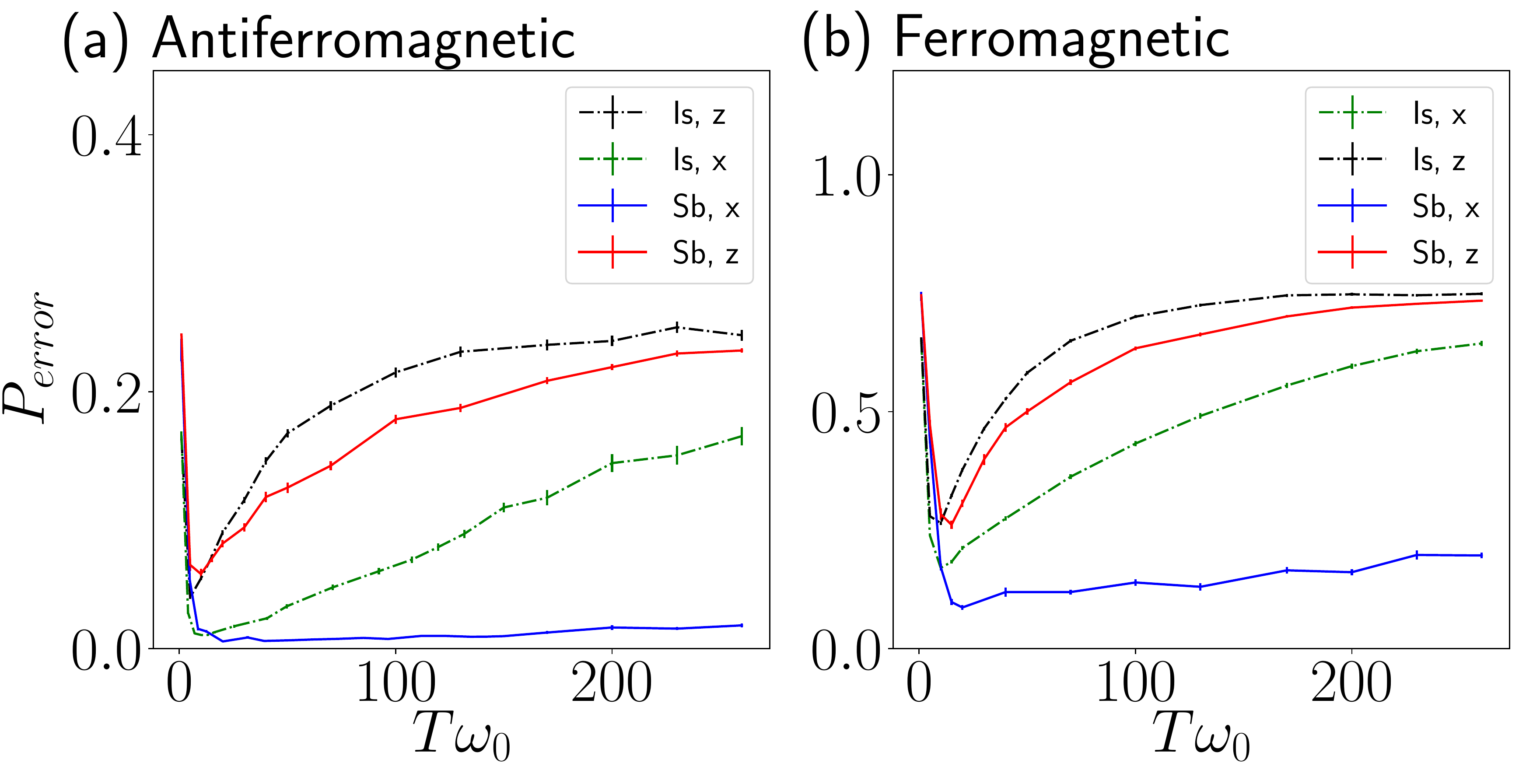}
\par\end{centering}
\vspace{0.1cm}
\caption{Error probability for QA as a function of total time of evolution $T$ for a system of $L=3$ spins with direct (Is) and with mediated couplings (SB). Annealing is performed for the Ising with transverse field model with  (a) ferro- and (b) antiferro-magnetic interactions.  The power spectrum of the noise is flat with strength $\gamma=0.1.$ A cutoff in the number of bosons per-site $n_{co}=8$ is used.
}\label{Fig:ann}
\end{figure}

To understand the combination of error suppression and correction, we have simulated an actual annealing passage using both ferro- and antiferro-magnetic couplings. Fig. \ref{Fig:ann} shows the error probability at the end of the passage, as a function of the total annealing time, for $L=3$ spins. The common feature of all the curves is a decrease of error probability at early times followed by an increase for intermediate times\ \cite{ashhab2006decoherence}, and a saturation at long times\ \footnote{Due to the strong white noise, at long times the system always saturates to a random spin configuration with $P_{error}(t\rightarrow \infty)=1-N_s/N,$ where $N_s/N$ is the fraction of ground state solutions over all spin configurations.}.  From Fig. \ref{Fig:ann}, it is clear that mediated couplings improves QA over the pure spin model for noise coupled in z and x direction. Similar to the curves in  Fig.\ \ref{Fig:1point}(c), we find that noise in the $z$ direction leads to a faster occupation of bosons.

\begin{figure}[t!]
\begin{centering}
\includegraphics[width=1\columnwidth]{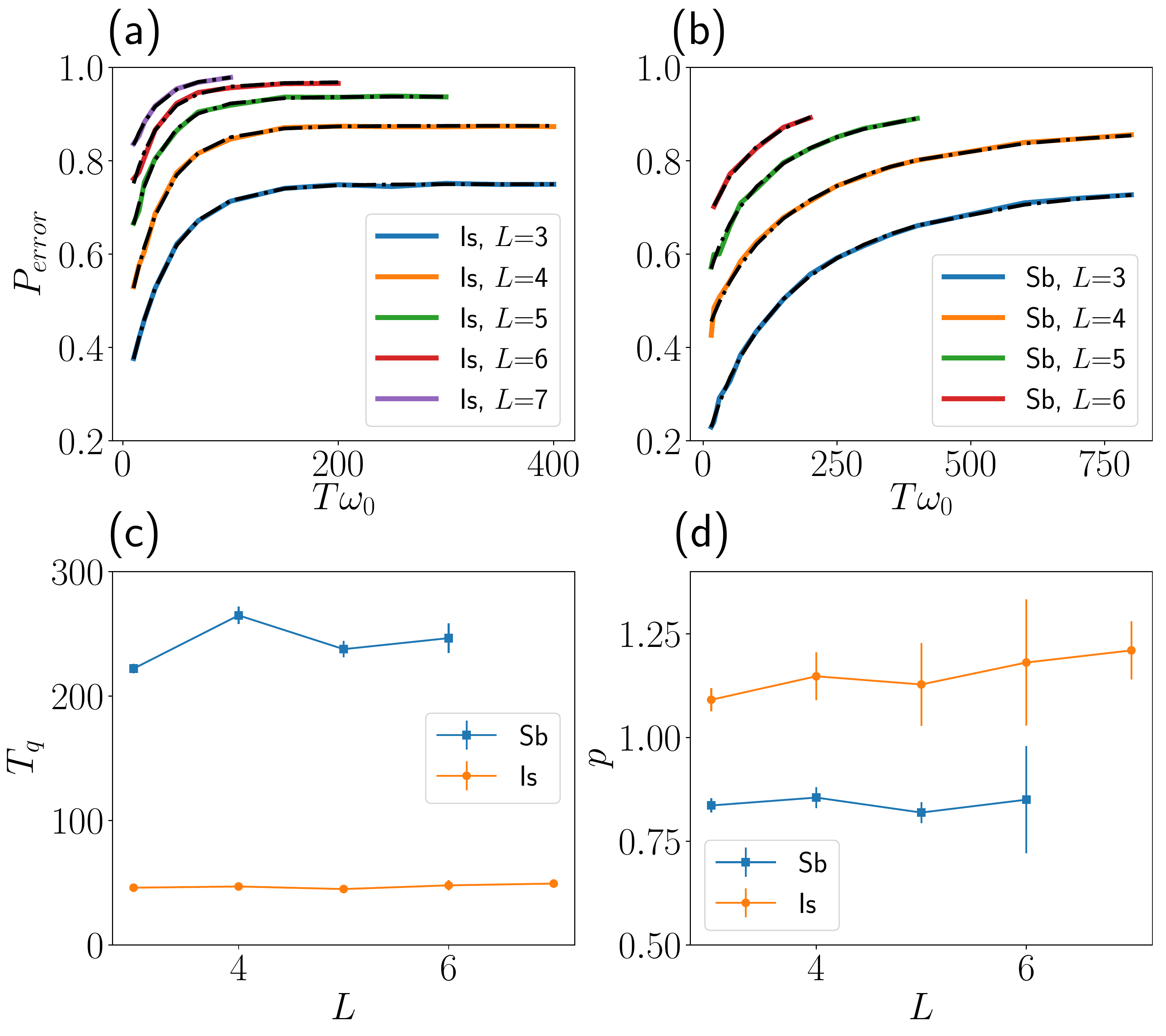}
\par\end{centering}
\vspace{0.1cm}
\caption{ Quantum annealing in the ferromagnetic Ising with transverse field for direct (a) and mediated (b) coupling and several sizes $L.$  The power spectrum of the noise is flat with strength $\gamma=0.2.$ The semi-dashed black  lines are fittings to the law $Y(T)=\{[1+a\ \exp(-(t/T_q)^p)]/2\}^L$ with free parameters $a,p,T_q.$ The parameters obtained in the fittings for $T_q$ and $p$ are plotted in (c) and (d), respectively.
A cutoff in the number of bosons per-site $n_{co}=4$ is used.
}\label{Fig_fseQA}
\end{figure}

The improvement of QA for mediated couplings and noise in $x$ direction is promising, but must be verified for larger sizes, analyzing finite size effects. Fig.\ \ref{Fig_fseQA} shows the error probability as a function of the total annealing time $T$ for (a) direct and (b) mediated couplings.  We have studied 3 to 7 qubits, using a similar number of bosonic modes in the SB, with a cut-off of $n_{co}=4$ excitations per mode. This allows computation of larger sizes, although it limits the heat bosonic modes can absorb.

The regions where $P_{error}$ grows with time can be fitted to the law from Eq.\ \eqref{Eq:fit}, obtaining both the qubit decay times $T_q$ and the decay power $p$ in panels (c) and (d), respectively. The Ising model gives a decay time $T_q\approx 50$ and $p\approx 1.2$ while for SB we obtain $T_q\approx 250$ and $p\approx 0.8.$ The goodness of the fitting improves significantly if values of $a<1$ are allowed in Eq.\ \eqref{Eq:fit}. This takes into account errors due to non-adiabatic transitions at small annealing times, and gives similar values for SB and Ising models. All of these implies a large extension of the decay time when interactions are mediated in the ultrastrong coupling regime. This increase is not an artifact of having a small number of qubits, as $p$ and $T_q$ do not show significant finite-size effects.

\paragraph{Discussion}

All our simulations show an asymmetry in the performance of the SB quantum annealer under errors, whereby $\sigma^x$ perturbations are more heavily suppressed and less harmful than $\sigma^z$ fluctuations. We can explain this result and argue that it scales to arbitrary sizes, by studying the SB Hamiltonian in the polaron basis. In the SB model, noise couples the ground state to excited spin and boson states $\ket{\widetilde{\psi}_{\alpha}^{(\textbf{n})}}=U^\dagger\ket{\varphi_{\alpha}^{(\textbf{n})}, \textbf{n}} $, via matrix elements of the form $\mathcal{M}_{\alpha,\textbf{n}}^{\theta}=|\braket{\varphi_\alpha^{(\textbf{n})},\textbf{n}|\widetilde{\sigma}_i^{\alpha}|\varphi_{gs}^{(\textbf{0})},\textbf{0}}|.$ These can be computed for the parameterization from Eqs.\ \eqref{eq:renor_1}, \eqref{eq:renor_2} using the spin operators in the polaron frame,   $\widetilde{\sigma}_i^{x}=\sigma_i^{x}$ and $\widetilde{\sigma}_i^{z}=\sigma_i^{z}\mathcal{U}^2.$ In the polaron transformed basis, $\mathcal{M}_{\alpha,\textbf{n}}^{x}\simeq 0$ for $\textbf{n}\neq 0$ and noise along $\sigma^x$ has a negligible probability to excite bosons.  Therefore, the increase in boson number in Fig.\ \ref{Fig:1point}(b) for noise in x direction is due to the mechanism of error correction, transforming spin mistakes into bosonic quasiparticles. Noise along $\sigma^z,$ on the other hand, couples to all bosonic configurations $\mathcal{M}_{q,\textbf{n}}^{z}\neq 0.$ This noise can heat the couplers without hybridization of spin and bosonic excitation, making it harder for the bosons to absorb or correct errors.

Another factor that explains the performance of errors is when they become relevant during the QA passage. Noise along $x$ and $z$  directions are more likely to create errors at the beginning and at the end of the passage, respectively. This means that the error correction mechanism is more efficient in mitigating the first type of noise, because errors that happen at later times have a lower probability to find the right avoided level crossing. This is also consistent with our numerical results in Fig.\ \ref{Fig:ann}, which shows a large decrease of error probability for annealing with mediated coupling under $x$ noise.

We can extract important conclusion for a realistic QA. It is reasonable to assume that the target Hamiltonian has a gap of the order of the effective qubit coupling $J(T)$ at the end of the passage\ \cite{altshuler2010anderson} and that the minimum gap along the passage is much smaller\ \cite{altshuler2010anderson,knysh2016zero,knysh2010relevance,pino2012,young2008size, mishra2018finite}. Then, one can design the couplers with frequency $\omega\lesssim J(T),$ so that avoided level crossings between low-energy excitations and bosonic modes would occur  after the minimum gap is attained. If thermal noise with small temperature is the main source of decoherence, errors are likely to occur due to low-energy excitations created around the minimum gap. This situation is similar to our computations with noise in $x$ direction because avoided level crossings take place after errors are introduced and noise cannot heat up the bosons for low enough temperature. As our results for noise in $x$, the error correction mechanism may well produce improvements of more than one order of magnitude in the effective qubit lifetime for realistic  annealers with mediated ultra-strong couplings.

In summary, we have provided strong evidences that the mechanisms of error reduction and error correction explained here could significantly reduce the effect of noise in intermediate-scale architectures of a QA. A first experimental test of our ideas should be possible with a few qubits devices that can be constructed with state-of-the-art technology in superconducting circuits\ \cite{yoshihara2017superconducting,kakuyanagi2016observation,forn2017ultrastrong,mukai2019superconducting}. We have also seen that the bosonic couplers improve the coherence of the quantum register, even when the Hamiltonian is not changed, as shown in Fig.\ \ref{Fig:1point}. This improvement in the information lifetime is due to the error suppression mechanism, which attenuates external fluctuations. The same idea can be used to improve the performance of other devices, such as quantum simulators, where one is interested in low-temperature dynamical properties\ \cite{hauke2016measuring}.

\begin{acknowledgments}
Financial support by Fundación General CISC (Programa Comfuturo) is acknowledged. J. J. García Ripoll acknowledges support from project PGC2018-094792-B-I00  (MCIU/AEI/FEDER, UE), CSIC Research Platform PTI-001, and CAM/FEDER Project No. S2018/TCS-4342 (QUITEMAD-CM). The numerical computations have been performed in the cluster Trueno of the CSIC.
\end{acknowledgments}

\bibliography{./spin_boson}

\pagebreak
\widetext

\setcounter{equation}{0}
\setcounter{figure}{0}
\setcounter{table}{0}
\setcounter{page}{1}
\makeatletter
\renewcommand{\theequation}{S\arabic{equation}}
\renewcommand{\thefigure}{S\arabic{figure}}
\renewcommand{\bibnumfmt}[1]{[S#1]}

\section*{Supplementary Material \--- Mediator assisted cooling in quantum annealing}

\section{ Minimum gap and complexity class}

We discuss the  minimum gap along an annealing passage for the Ising model in transverse field when qubit interactions are direct, Eq. (1) of the main text, and mediated, Eq. (2).  In the case of direct couplings, it is well known that the minimum gap closes as $\Delta_m\sim L^{-z}$,  with number of qubits $L$ and dynamical critical exponent $z=1.$  It was shown numerically and with the polaron ansazt that the SB in the ultra-strong coupling regime is within the same universality class\ \cite{kurcz2014hybrid}, so the former scaling law applies with equal dynamical critical exponent. This implies that the complexity class of annealing is the same for direct and mediated interactions, as adiabaticity roughly imposes $T\gg\Delta_m^{-2}$ with $T$ total annealing time. Here, we go further and check that the constant that enters in the scaling law  are of the same order for both models. 

\begin{figure}[b!]
\begin{centering}
\includegraphics[width=0.5\columnwidth]{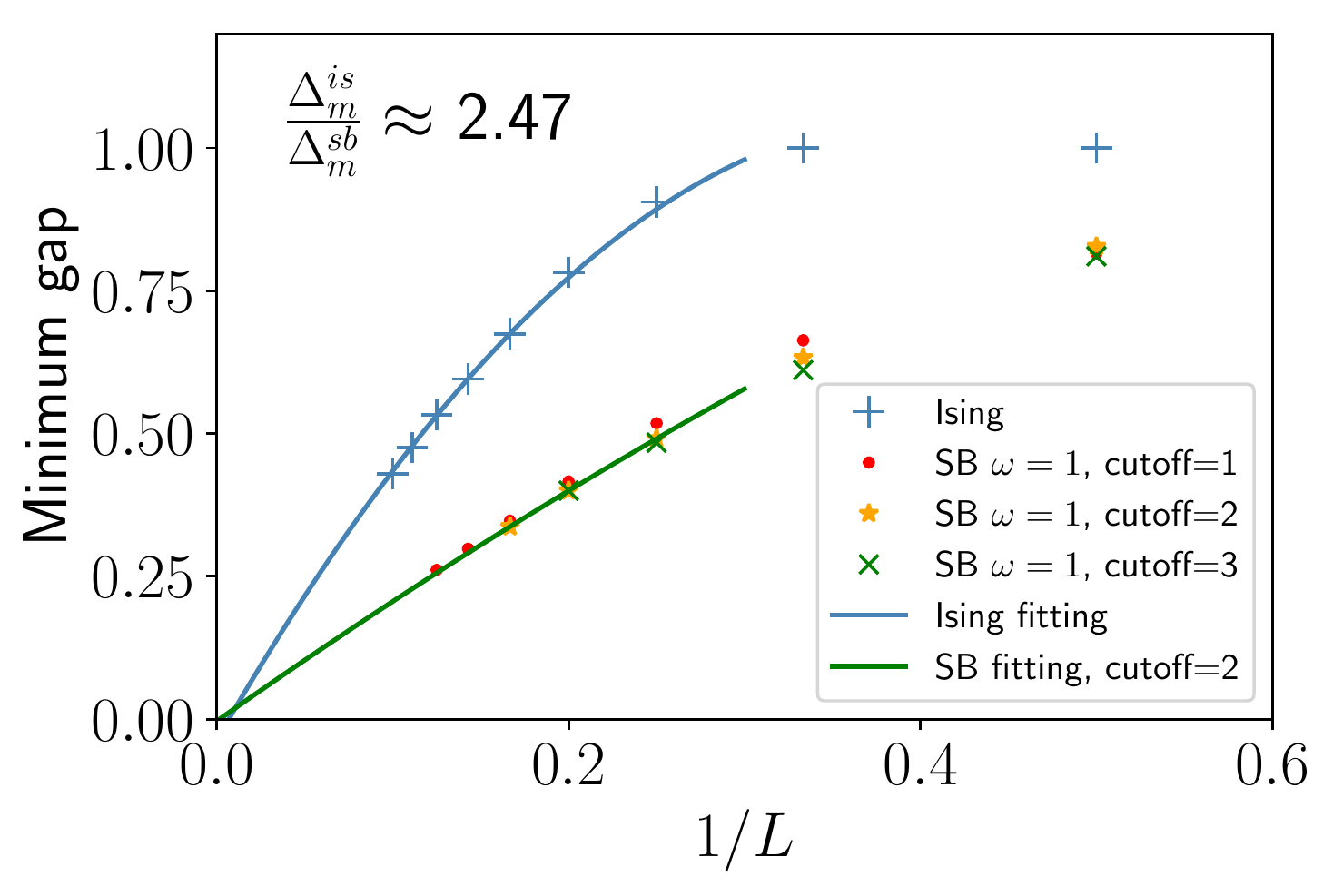}
\par\end{centering}
\vspace{0.1cm}
\caption{ Minimum gap as a function of number of spins, $L$, for annealing in the Ising model with transverse field. The symbols $+$ corresponds to direct interactions implemented with an Ising-like Hamiltonian, while symbols $\bullet, \star, \times$ are used for SB model in the ultra-strong coupling regime with different cutoffs in the number of bosonic excitations allowed in each resonator. The solid lines 
are fittings to a law $a*L^{-1}+b*L^{-2}$ for the Ising and SB with cutoff equal to two. The results gives for Ising model $a = 5.36$, $b =-0.36$ and for SB $a=2.17$, $b=-0.0034.$
}\label{Fig:mgap}
\end{figure}

According to previous paragraph, the minimum gap along an annealing passage closes as $\Delta_m =a L^{-z}$, with $z=1.$ We compare the proportionality constant $a$ for direct and mediated couplings.  In Fig\ \ref{Fig:mgap}, we have plotted the minimum gap for Ising-like and SB Hamiltonians.  In the latter case, three different cutoffs in the number of bosons have been used. The curves for different cutoffs are quite similar, so we use the largest cutoff data as a good approximation of the full SB model. We have fitted the minimum gap of the Ising and SB to a law $\Delta_m= a L^{-1}+bL^{-2}$, with free parameters $a,b.$   The parameters $b$ is introduced to allow for irrelevant corrections.  The result of the fittings gives constant for Ising-like and SB that are related by $a_{is}\approx 2.5 a_{sb}.$

In summary, we have found that not only the complexity class is the same but also the constant that  enters in the scaling law of the minimum gap are of the same order.  Thus, the amount of errors induced by non-adiabatic transition for QA  are very similar for architectures with direct and mediated couplings.

\section{ Numerical implementation of noise}\label{App:noise1q}

We now provide details about how we have implemented the noise in our numerical simulations. First, we introduce the definitions for a single qubit:
\begin{equation}
H = \frac{1}{2}\left[\omega_0\sigma^z+f(t)   \sigma^\theta\right],\label{Eq:1q}
\end{equation}
where $\omega_0$ is the qubit frequency and the function $f(t)$ is a random process that represent the noise acting on the qubit. This noise couple with the qubit in a direction with angle $\theta$ respect the $z$-axes: $\sigma^\theta = \cos(\theta)\sigma^z +\sin(\theta)\sigma^x.$  Dephasing and relaxation times depend on angle $\theta$ and power spectrum. One way to define these times is via the decay of the initial qubit state after averaging over  different realizations of noise:
\begin{align}
\overline{\av{+_z|\sigma^z(t)|+_z}}&= \exp{(-t/T_1)}\\
\overline{\av{+_x|\sigma^x(t)|+_x}}&= \cos(\omega_0 t) \exp{(-t/T_2)}
\end{align}
Average over disorder realization is  denoted by a line over the quantity to average. The values of these times depends on the angle of coupling 
between the qubit and noise and on the power spectrum of the noise\ \cite{paladino20141}:
\begin{align}\label{eq:th}
\frac{1}{T_1} &= \pi\sin^2(\theta) S(\omega_0)\\
\frac{1}{T_2^\star}&=\pi \cos^2(\theta)S(\omega\rightarrow 0)\\
\frac{1}{T_2}& = \frac{1}{T_2^\star}+\frac{1}{2T_1}.            
\end{align}
The auto-correlation function and power spectrum of the noise, that appears in previous formulas, are defined as:
\begin{align}
 \hat{S}(\tau)&=\av{f(t+\tau)f(t)},\\
 S(\omega) &=\frac{1}{2\pi}\int_{-\infty}^{\infty} S(\tau) e^{-i\omega \tau}d\tau,
\end{align}
where the notation $\av{\dots}$ means time average.  

\begin{figure}[t!]
\begin{centering}
\includegraphics[width=0.5\columnwidth]{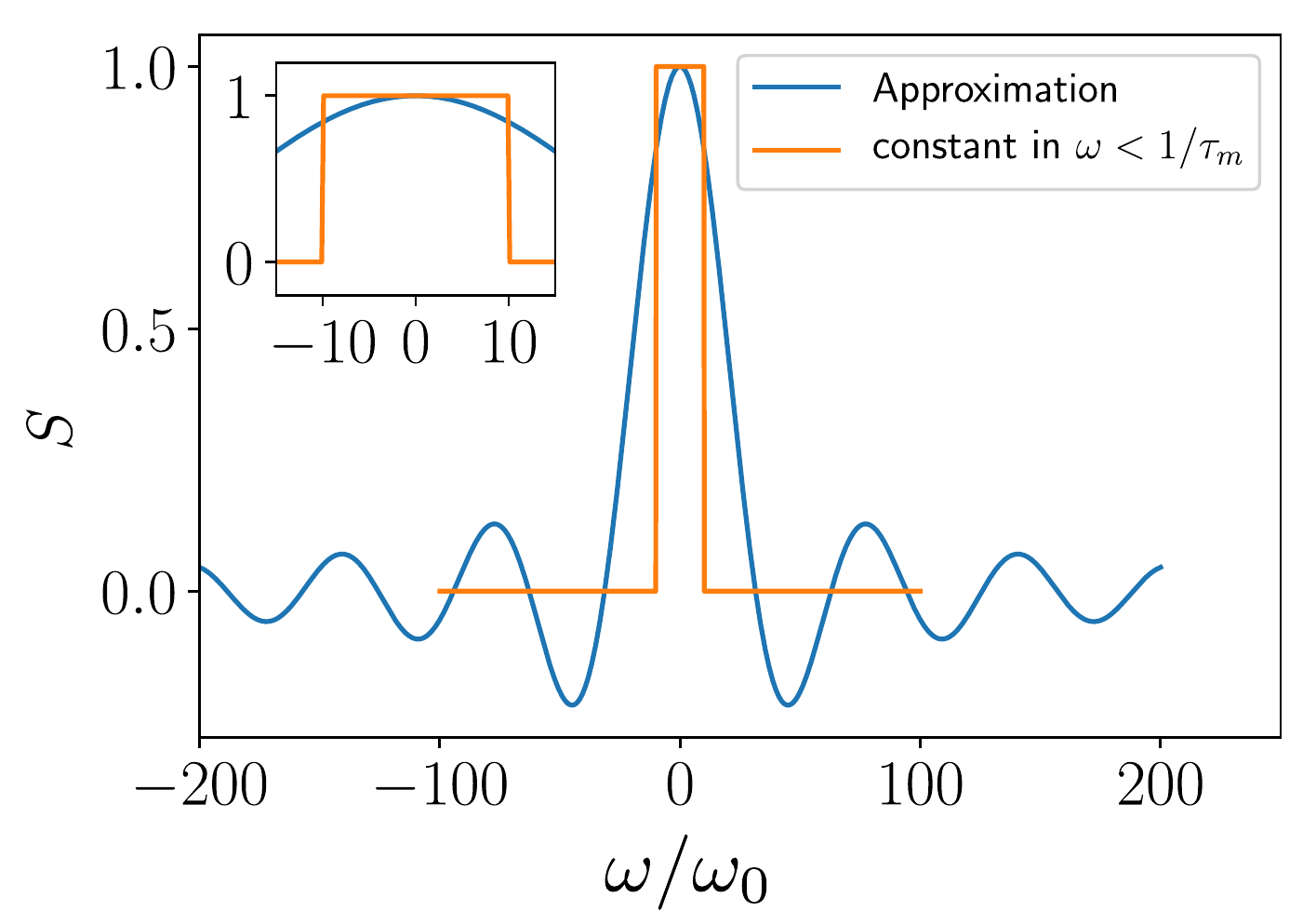}
\par\end{centering}
\vspace{0.1cm}
\caption{ The power spectrum described in equation \eqref{eq:ps_wnn} which corresponds to a noise with auto-correlations as in equation \eqref{eq:ac_wnn}. The values of parameter are $\gamma=2\pi$ and $\tau_m=0.1/\omega_0$, being $\omega_0$ qubit frequency. In the inset we can see a zoom to the region of small frequencies. 
}\label{Fig_ls}
\end{figure}

We use Gaussian noise, so distribution probability is at any time:
\begin{align}
P[f(t) ]= \frac{e^{-\frac{[f(t)]^2}{2\gamma^2}}}{\sqrt{2\pi\gamma^2}}.
\end{align}
Furthermore, we work with noise that is correlated only on small times. Theoretically, one can define white noise with the property of:
\begin{equation}
 \av{f(t)f(t+\tau)}=\gamma^2 \delta(\tau\omega_0).
\end{equation}
The  power spectrum is easy to compute using the previous equation:
\begin{align}
S(\omega)&=\frac{1}{2\pi}\frac{\gamma^2}{\omega_0}.
\end{align}
The formulas for decoherence and relaxation times are for white noise:
\begin{align}
\frac{1}{T_1} &=  \sin^2(\theta)\  \frac{\gamma^2}{2\omega_0}\label{eq:th1}\\
\frac{1}{T_2^\star}&= \cos^2(\theta)\ \frac{\gamma^2}{2\omega_0}\label{eq:th2}
\end{align}

\subsubsection{Approximation of white noise}

White noise is a theoretical idealization that involves arbitrary large frequencies which  cannot be reproduced in numerical simulations.  Here, we use uncorrelated noise for the typical frequencies of the qubit $\omega_0$ but correlated for much larger frequencies $\omega\gg\omega_0.$ This can be done by splitting a time interval $T$ in $N$ subintervals and set a noise:
\begin{equation}
f(t)=  \Big\{  f_i\sqrt{\frac{\gamma^2}{\tau_m\omega_0}} \quad\text{if}   \quad i*\tau_m<t<(i+1)*\tau_m \Big\}\label{eq:ac_wnn}
\end{equation}
where $\tau_m=T/N\ll 1$ and $t$ run from 0 to T. The $f_i$ are uncorrelated dimensionless numbers following a Gaussian distribution with zero mean and variance 1. The auto-correlation for this type of noise is:
\begin{equation}
 \hat{S}(\tau)=\begin{cases}
       0 &\quad\text{if}\ \ \ \tau> \tau_m\\
       \frac{\gamma^2}{\omega_0\ \tau_m} &\quad\text{if}\ \ \ \tau< \tau_m \\ 
     \end{cases}
\end{equation}
Power spectrum is then:
\begin{equation}
 S(\omega) = \frac{1}{2\pi}\Re \big\{ \int_{-\infty}^{\infty} e^{-i\omega\tau}\hat{S}(\tau)d\tau \big\}= \frac{1}{2\pi}\Re \big\{ \frac{\gamma^2}{\omega_0\tau_m} \int_{0}^{\tau_m}  e^{-i\omega\tau}d\tau\big\} = \frac{\gamma^2}{2\pi\omega_0}\left[\frac{\sin(\tau_m\omega)}{\tau_m\omega}\right] \label{eq:ps_wnn}
\end{equation}
This noise has an approximated constant power for frequencies $\omega<1/\tau_m$, while it oscillates for larger ones. The power spectrum of the noise and a square function with width given by $\tau_m\omega_0=0.1$ are plotted in Fig.\ \ref{Fig_ls}. Setting $\tau_m\ll 1/\omega_0$ allows to have a pretty flat spectrum for the relevant frequencies when simulating a system with typical frequency given by $\omega_0.$

\begin{figure}[t!]
\begin{centering}
\includegraphics[width=0.8\columnwidth]{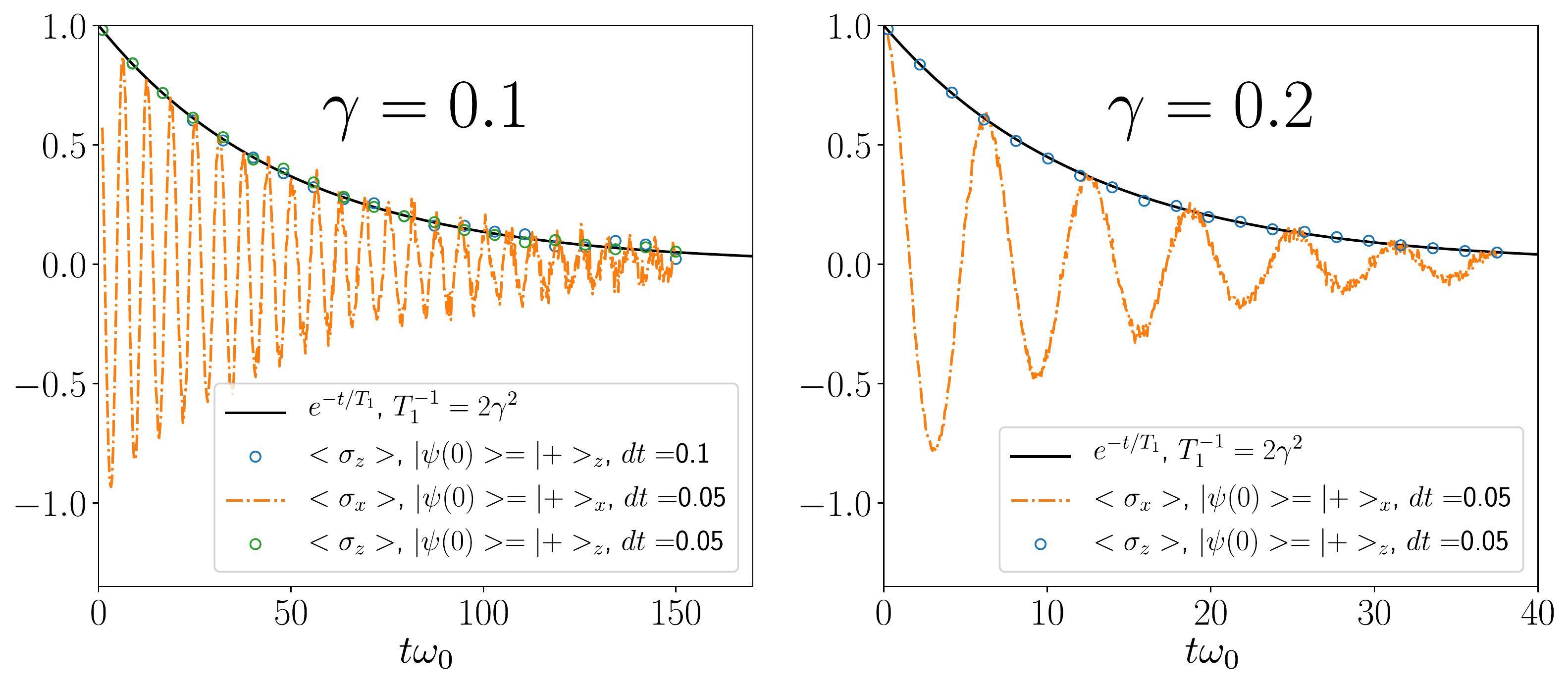}
\par\end{centering}
\vspace{0.1cm}
\caption{ Time evolution under noise of $\av{\sigma^x}$ (semi-dotted) and $\av{\sigma^z}$ (bullets) of one qubit with initial states in $+x$ and in $+z$ directions.   Noise is coupled to the qubit in directions $\theta = 0$ and $\theta= \pi/2$ for each case.  Results for two different time steps in the Lanczos method , $dt=0.1$ in green and $dt=0.05$ in blue, give the same  decay for $\av{\sigma^z}.$ The theoretical predictions Eq.\ \eqref{eq:th1} and\ \eqref{eq:th2}  for noise in $z$ direction are represented with black solid curves. Each panel correspond to different values of $\gamma$ which are $0.1$ (left) and $0.2$ (right). Averages have been performed over, approximately, $5000$ realizations of the noisy signal.
}\label{Fig_1qnoise}
\end{figure}

\subsubsection{Numerical method}

The numerical simulations of a system of $L$ qubits have been performed using $L$ uncorrelated random process as the one in Eq.\ \eqref{eq:ac_wnn}. Each of these functions represent a local noise, so the total Hamiltonian is:
\begin{equation}
    H= H_0 + \frac{1}{2}\sum_i f_i(t)\sigma_i^{\theta}\label{Eq:full},
\end{equation}
where noise couples with all the qubits in the same angle $\theta.$ The Hamiltonian $H_0$ contains the dynamic of the noise-free system. 

We have used Lanczos method for numerical simulation of  time evolution of Hamiltonians as in Eq. \eqref{Eq:full} for the results presented in the main body of the work. This method can cope with time dependent Hamiltonian, as it is needed in quantum annealing, but one has to approximate the Hamiltonian as a constant during an integration time step $dt.$ For the simulation of noise, this implies that Lanczos method imposes the high frequency cutoff in the power spectrum of the noise at $\omega_h=1/dt.$ We have then chosen a correlated noise with $\tau_m=dt$, as described in Eq.\ \eqref{eq:ac_wnn}, and set $dt=0.1\omega_0.$ We  checked that simulations for $dt=0.1$  and $dt=0.05$ give the same results.

Finally, let us discuss the results of time evolution for the Hamiltonian of one-single qubit, Eq. \eqref{Eq:1q}, using our method. We have studied the cases of pure dephasing $\theta=0$ and relaxation $\theta=\pi/2.$ The results for $\gamma=0.1, 0.2$ appears in Fig.\ \ref{Fig_1qnoise}. Simulations with two times steps in the Lanczos algorithm for  $\theta=\pi/2$ and $\gamma=0.1$ appear with green and blue circumferences and show no differences as expected. 
The solid lines represent the decay following laws Eqs.\ \eqref{eq:th1} and \eqref{eq:th2}. The agreement between numerical curves and theory is excellent.

\end{document}